# Exploring Multi-Objective Trade-offs in Reference Compound Selection for Validation Studies of Toxicity Assays

Yohei Ohto[1], Yasuhiro Yoshikai[1], Hiromi Fujimoto[1], Kaoru Sato[2], Hiroyuki Kusuhara[1], Tadahaya Mizuno[1, 3†]

1 Laboratory of Molecular Pharmacokinetics, Graduate School of Pharmaceutical Sciences, The University of Tokyo, 7-3-1 Hongo, Bunkyo, Tokyo, Japan
2 National Institute of Health Sciences, 3-25-26 Tonomachi, Kawasaki-ku, Kawasaki City, Kanagawa Prefecture, Japan
3 The Institute of Statistical Mathematics (ISM), Research Organization of Information and Systems, Tachikawa, Tokyo, 190-8562, Japan.
†Author to whom correspondence should be addressed

## Abstract

In chemical safety assessment, validation studies rely on reference compound lists to evaluate the applicability of alternative methods prior to regulatory acceptance. These lists are expected to cover multiple aspects, including chemical structure, physicochemical properties, and toxicity profiles. In practice, however, trade-offs among these aspects are typically addressed implicitly through expert judgment, making them difficult to examine systematically. Here, we formulate reference compound selection for toxicity assay validation as an explicit multi-objective design problem. We define three interpretable objectives capturing structural, physicochemical, and toxicity diversity, and employ a genetic algorithm as an exploratory tool to examine the trade-off structure and resulting Pareto-optimal solutions. Rather than prescribing optimal or recommended compound sets, this formulation enables systematic exploration of designs and explicit comparison of their positions within a common design space. As an illustrative application, we link existing assay datasets with expert-curated validation lists by treating "selected as a reference compound" as an annotation on the underlying compound pool. We show that expert-selected, random, and algorithmically generated compound lists occupy distinct regions of the design space. Furthermore, under an illustrative fixed modeling setup, different regions of the design space were associated with different observed evaluation outcomes, supporting the view that reference compound selection constitutes a structured dimension of evaluation design. Together, these results provide a methodological perspective for treating reference compound selection as an analyzable design object, complementing established expert-driven practices.

# 1 Introduction

In chemical safety assessment, the evaluation of potential hazards posed by chemical substances is essential[1]. Animal testing has long played a central role in safety assessment; however, ethical concerns, high costs, long testing periods, and interspecies variability have motivated the development of alternative approaches[2,3]. In response, a range of alternative safety assessment methods, including in silico prediction models[4,5] and in vitro toxicity assays[6–8], have attracted increasing attention.

Before regulatory acceptance, these methods must undergo validation to assess their predictive performance, reliability, and reproducibility[9–11]. Validation studies of toxicity assays typically rely on carefully curated reference compound lists, which are designed to cover appropriate ranges of toxicity potency, chemical structure, and physicochemical properties[10]. These reference compounds are generally selected by domain experts, informed by historical precedent, experimental feasibility, and available toxicity databases, often in alignment with regulatory frameworks such as the Globally Harmonized System of Classification and Labelling of Chemicals (GHS)[12,13].

Despite their central role, reference compound lists are rarely treated as explicit objects of methodological design. Coverage along one dimension—for example, toxicity potency—may constrain or trade off against coverage along others, such as structural or physicochemical diversity. In practice, these trade-offs are addressed through expert judgment, which is both necessary and effective, but the resulting trade-off structure remains implicit, making it difficult to analyze and compare systematically. For example, in assay-specific toxicity validation, emphasizing diversity in toxicological responses may narrow the range of chemical scaffolds represented in the set, whereas prioritizing broad structural coverage can introduce compounds that are less directly aligned with the assay context or endpoint-specific relevance criteria.

From a methodological perspective, reference compound selection can be viewed as a design problem defined under multiple, potentially conflicting objectives. Different selections can occupy distinct positions within the design space, even when the size of the compound list and the validation protocol are held fixed. Making these trade-offs explicit provides a transparent analytical basis for discussing how design choices shape validation datasets.

In this study, we formulate reference compound selection for validation of toxicity assays as a multi-objective trade-off problem. Rather than prescribing optimal compound sets, we use a genetic algorithm (GA) as an exploratory tool to examine the structure of this trade-off space[14]. GA is well suited for this purpose because it can efficiently explore complex combinatorial spaces and identify Pareto-optimal solutions, where improvement along one objective necessarily involves compromise along others[15]. The primary contribution of this study lies not in proposing a new optimization algorithm or a preferred reference list, but in formalizing reference compound selection as an explicit and comparable design component in toxicity assay validation. Accordingly, the present study should be read as a methodological case study on validation design, rather than as a benchmark comparison of predictive models or a prescriptive framework for selecting recommended compound sets.

# 2 Related Work

## 2.1 Multi-Objective Optimization and Genetic Algorithms

Multi-objective optimization provides a natural framework for problems in which several desirable properties must be considered simultaneously under inherent trade-offs. In such settings, improvement in one objective often requires compromise in another, making Pareto-optimal solutions a useful representation of achievable design alternatives. Genetic algorithms (GA) have been widely used to explore such trade-off structures in complex combinatorial spaces, particularly when objective functions are nonlinear or discontinuous[15–18].

In the present study, GA is not introduced as a methodological contribution in itself. Rather, it is used as an exploratory tool for examining feasible regions of the design space and visualizing the Pareto structure induced by multiple design objectives. This role is consistent with the aim of the study, which is to formalize reference compound selection as an analyzable design problem rather than to propose a new optimization algorithm or to identify a uniquely preferred solution.

### 2.2 Evaluation Design and Dataset Construction in Cheminformatics and Computational Toxicology

In cheminformatics and computational toxicology, evaluation outcomes depend strongly on how datasets are constructed and partitioned. Prior studies have shown that commonly used random-split validation can yield results that differ substantially from those obtained under more application-oriented settings, such as time-based splits, which better reflect temporal distribution shifts encountered in practice[19,20]. This line of work has established evaluation design as an important methodological issue in predictive modeling.

A representative example is Simulated Medicinal Chemistry Project Data (SIMPD), which uses a multi-objective GA framework to generate training–test splits that emulate temporal progression in medicinal chemistry projects[21]. Such studies address how data partitioning affects the realism of model evaluation[22]. The present study is related in spirit but addresses a different question. Rather than focusing on how an existing dataset should be split, we examine how the internal composition of a reference compound set can itself be treated as an explicit design variable in validation studies. In this sense, reference compound selection and data-splitting strategies operate at different but complementary levels of evaluation design.

This distinction is particularly relevant in toxicity assay validation, where reference compound lists are often expert-curated and assay-specific, yet the trade-off structure underlying their composition is rarely made explicit. Our work therefore builds on prior evaluation design studies while shifting attention from partition strategy to reference-set composition as a comparable object of methodological analysis.

## 3 Methods

### 3.1 Validation Studies and Data Sources

#### 3.1.1 Validation Studies

The present study analyzed reference compound lists officially used in validation studies conducted by the Japanese Center for the Validation of Alternative Methods (JaCVAM)[23] as case-study datasets for methodological exploration. These validation studies cover acute, endocrine, and developmental toxicity assays that have been internationally evaluated and are aligned with corresponding OECD Test Guidelines.

The selected assays span multiple toxicological endpoints and application contexts, reflecting the diversity of chemical safety assessment practices rather than a focus on a specific regulatory domain or chemical class. In particular, for endocrine disruption, multiple assays targeting estrogen and androgen receptor signaling—VM7Luc ER TA, ER STTA, AR-EcoScreen, AR-CALUX, and hrER binding—were included to capture distinct endpoints such as receptor transactivation and ligand binding.

Full assay specifications, numbers of reference compounds, and links to external toxicity databases are provided in **Table 1** and **Supplementary Table 1**. In this study, these validation studies are treated as fixed design contexts, and no attempt is made to reinterpret their regulatory conclusions or experimental outcomes.

### 3.1.2 Large-Scale Toxicity Databases

To link the reference compounds from JaCVAM validation studies with publicly available toxicity information, three large-scale databases— ICE[24], TDC[25], and Tox21[26]—were employed (see **Table 1** and **Supplementary Table 1**). Each database covers complementary aspects of chemical and biological variation, allowing consistent extraction of physicochemical and toxicity data across assays.

The Integrated Chemical Environment (ICE), developed by the National Toxicology Program Interagency Center for the Evaluation of Alternative Toxicological Methods (NICEATM, U.S. NTP), provides curated datasets integrating in vitro, in vivo, and in silico toxicity data. In this study, ICE served as the primary source for acute oral toxicity and developmental toxicity information. Acute toxicity data included quantitative measures such as $LD_{50}$ values, whereas developmental toxicity data were obtained from the DART (Developmental and Reproductive Toxicology) dataset, which includes endpoints such as differentiation inhibition indices.

The Therapeutics Data Commons (TDC) is an open-source platform integrating multiple biomedical resources, including Tox21[26], SIDER[27], and ChEMBL. In this study, TDC was used as the primary access point for endocrine disruption–related assays, including VM7Luc ER TA, ER STTA, AR-EcoScreen, AR-CALUX, and hrER binding. For each compound, corresponding assay endpoints—such as receptor transactivation activity or binding affinity—were retrieved using standardized SMILES-based identifiers provided through the TDC interface.

Tox21 is a joint initiative of the U.S. EPA, NIH, and FDA that provides high-throughput screening (HTS) data aimed at reducing reliance on traditional animal testing. In this study, Tox21 data were accessed via the TDC interface to supplement receptor-related assay endpoints for estrogen and androgen receptor signaling. The broad assay coverage of Tox21 enabled inclusion of compounds spanning diverse biological response profiles, supporting systematic exploration of reference compound selection across multiple assay contexts.

For each compound included in the JaCVAM validation studies listed in **Table 1**, the corresponding CAS Registry Number (CAS-RN) was used to retrieve the canonical SMILES representation via the PubChem API[28]. All SMILES were standardized and converted into isomeric SMILES using RDKit[29] (version 2024.09.5), ensuring consistent structural encoding across databases.

Using these standardized identifiers, compounds were matched to entries in the databases described above, thereby defining a candidate compound pool for each validation study. Acute and developmental toxicity assays were primarily linked to ICE datasets, whereas endocrine disruption assays were mapped to corresponding endpoints available through TDC and Tox21.

When an isomeric SMILES match was available, toxicity data were adopted directly from the database. When database entries were unavailable, toxicity information was extracted from the original JaCVAM validation documents (**Supplementary Table 1**). Differences in experimental protocols, endpoint definitions, and reporting conventions across assays and databases were not reconciled. Accordingly, toxicity values and labels were treated as assay-specific descriptors rather than as directly comparable endpoints across assays.

For regression-type endpoints (e.g., $LD_{50}$), numerical values were normalized to unit variance. For classification-type endpoints (e.g., agonist versus antagonist), binary labels followed the conventions of the source database. These data integration and preprocessing steps were applied uniformly across all analyses.

## 3.2 Problem Formulation

### 3.2.1 Reference Compound Selection as a Design Problem

In this study, reference compound selection for toxicity assay validation is formulated as an explicit design problem. Given a fixed validation context—defined by assay type, endpoint definition, and compound list size—the task is to select a subset of compounds from a larger candidate compound pool such that multiple, potentially conflicting design objectives are jointly considered.

Importantly, this formulation does not aim to identify an optimal or recommended reference compound list. Instead, it provides a structured formulation for exploring how different design choices position compound lists within the design space defined by the specified objectives. Expert-driven reference lists used in existing validation studies are therefore treated as one realization within this design space, rather than as a ground truth or target solution.

### 3.2.2 Decision Variables and Constraints

The decision variable in this formulation is the composition of a reference compound list of fixed size $n$, selected from a candidate compound pool of compounds associated with a given validation study. The list size $n$ is constrained to match that of the corresponding JaCVAM validation study to enable direct comparison under consistent experimental contexts.

Additional constraints are imposed implicitly by data availability: only compounds for which chemical structures and assay-specific toxicity information can be retrieved from public databases or original validation documents are considered. No constraints related to compound cost, availability, or experimental feasibility are included, as such information is not consistently available in structured form.

### 3.2.3 Objective Functions

To explore the design space of reference compound selection, three objective functions were defined to capture complementary aspects that are commonly considered in the construction of reference compound lists for toxicity assay validation: chemical structure diversity, physicochemical property diversity, and toxicity diversity.

These objectives do not represent regulatory definitions or exhaustive criteria for reference compound selection. Rather, they are operational proxies introduced to enable systematic and transparent exploration of trade-offs under explicit and controlled assumptions. Alternative formulations or additional objectives could be adopted depending on the specific validation context; the present study deliberately restricts the objectives to these three dimensions to maintain interpretability and computational tractability.

Chemical structure diversity reflects the extent to which compounds differ in their molecular scaffolds and substructures, and is commonly emphasized to ensure broad coverage of chemical space. Physicochemical property diversity captures variation in properties relevant to assay behavior and compound disposition, such as lipophilicity and polarity. Toxicity diversity represents variation in assay-specific response profiles, such as potency ranges or class balance, which is important to avoid overrepresentation of narrow response regimes in validation datasets.

For all objectives, diversity was quantified at the level of compound pairs within each reference compound list. The objective functions were formulated such that lower overall similarity or greater dispersion corresponds to higher diversity. These formulations allow different reference compound lists of fixed size to be compared within a common multi-dimensional objective space, enabling explicit visualization of trade-offs among competing design criteria.

The specific mathematical definitions of each objective function are provided in the following sections. Importantly, the resulting objective values should be interpreted as relative indicators within the defined design space, rather than as absolute measures of reference compound quality.

### 3.3 Multi-Objective Optimization Framework

### 3.3.1 Genetic Algorithm Implementation

To explore the multi-objective trade-off structure defined in Section 3.2, we employed a GA as an exploratory computational approach. GA was chosen not to identify an optimal or recommended reference compound list, but to efficiently sample and visualize feasible regions of the design space under multiple competing objectives.

The optimization was implemented using a standard NSGA-II (Non-dominated Sorting Genetic Algorithm II)[30]–based multi-objective genetic algorithm. Candidate reference compound lists of fixed size were represented as individuals, and evolutionary operations including selection, crossover, and mutation were applied iteratively to generate diverse compound lists spanning the objective space.

At each generation, individuals were evaluated according to the defined objective functions, and non-dominated sorting was used to identify Pareto-optimal solutions. Crowding distance was employed to maintain diversity along the Pareto front, ensuring coverage of distinct regions of the design space rather than convergence to a single solution.

The final output of the optimization is the Pareto front, representing the boundary of feasible reference compound list designs under the specified objectives and constraints. Importantly, the structure of the Pareto front is independent of any weighting among objectives. All subsequent analyses therefore refer explicitly to the distribution and structure of solutions on the Pareto front, rather than to a single optimized list. Implementation details, including population size, number of generations, crossover and mutation rates, and software versions, are provided in the **Supplementary Notes**.

### 3.3.2 Diversity Metrics

To operationalize the objectives defined in Section 3.2.3, three diversity metrics were used to quantify complementary aspects of reference compound list composition: structural diversity, physicochemical property diversity, and toxicity diversity. All metrics were computed at the level of compound pairs within each reference compound list and formulated such that the objective functions consistently reflect diversity, either by minimizing similarity or maximizing dispersion.

Structural diversity was quantified using pairwise similarity between molecular fingerprints. For each pair of compounds within a list, the Tanimoto similarity of their ECFP4 fingerprints[31] was calculated, and the sum of pairwise similarities was minimized:

$$\sum_{i=1}^{n}\sum_{k=1}^{i-1} Tanimoto\ similarity(ECFP(c_i),\ ECFP(c_k)) \rightarrow min.$$

Minimizing this quantity corresponds to reducing structural redundancy within the list and increasing coverage of distinct chemical scaffolds.

Physicochemical property diversity was quantified based on dispersion in a low-dimensional property space. For each compound, the LogP value[32] and the TPSA value[33] were calculated, standardized using a robust z-score[34], and pairwise Euclidean distances were computed:

$$\sum_{i=1}^{n}\sum_{k=1}^{i-1} D(c_i, c_k) \rightarrow max$$

$$D(c_i, c_k) = \left\| \begin{pmatrix} robust_z(logp(c_i)) - robust_z(logp(c_k)) \\ robust_z(TPSA(c_i)) - robust_z(TPSA(c_k)) \end{pmatrix} \right\|_2 .$$

The sum of pairwise distances was maximized, such that larger values indicate broader dispersion of physicochemical properties across the compound list.

Toxicity diversity was evaluated in an assay-specific manner. For continuous toxicity endpoints (e.g., $LD_{50}$), diversity was quantified using the coefficient of variation of log-transformed toxicity values within the list. For binary endpoints (e.g., toxic versus non-toxic, agonist versus antagonist), diversity was quantified using the log-likelihood of a binomial distribution with fixed success probability of 0.5, which assigns higher scores to more balanced class compositions. Because toxicity endpoints differ across assays in scale and type (continuous or binary), the corresponding mathematical definitions are provided in the **Supplementary Notes**.

These metrics are not intended as definitive or regulatory measures of diversity. Rather, they serve as interpretable operational proxies that enable systematic comparison of reference compound lists within a common multi-objective design space. Additional details regarding metric definitions, normalization procedures, and alternative formulations are provided in the **Supplementary Notes**.

### 3.4 Illustrative Evaluation Analyses

To illustrate how reference compound selection shapes downstream evaluation characteristics, we conducted two complementary analyses based on the compound lists generated by the multi-objective optimization. These analyses are intended as illustrative assessments of the implications of different design choices, rather than as benchmarks for comparing predictive models. For visualization and illustrative comparison purposes only, a single illustrative Pareto-optimal example was selected from the Pareto front using a composite score; details are provided in the **Supplementary Notes**.

First, compound lists generated by the GA were compared with randomly generated lists and the original reference lists used in validation studies using the same objectives defined in Section 3.3.2. This comparison characterizes how different selection strategies position reference compound lists within the multi-objective design space defined by structural, physicochemical, and toxicity diversity.

Second, evaluation outcomes were examined using a representative machine learning model for toxicity prediction. For each reference compound list, models were trained on the remaining compounds and evaluated on the selected list. Differences in predictive performance metrics (e.g., accuracy and AUC) were interpreted as reflecting differences in test set composition and heterogeneity, rather than intrinsic differences in model capability.

These analyses do not aim to establish the superiority, rigor, or preference of any particular evaluation setup. Instead, they are intended to show that, under otherwise comparable modeling conditions, reference compound selection constitutes a structured dimension of evaluation design and contributes to variation in observed evaluation outcomes. Implementation details of the modeling and evaluation procedures are provided in the **Supplementary Notes**.

### 3.5 Data and Code Availability

The data, code, and results generated in this study are available in the following GitHub repository: https://github.com/mizuno-group/multi-objective-optimization.

## 4 Results

In this section, we report the observed characteristics of reference compound lists and their distributions within the design space defined in Section 3. The present study is intended as a methodological case study across several validation contexts rather than as a comprehensive survey of all toxicity assay types. Results are presented primarily using Validation Assay 09_02, specifically the ER-STTA assay for detecting antagonist activity in endocrine disruption screening[35], as a representative case. Results for the 09_02 agonist assay and Validation Assay 07_02 (3T3 Neutral Red Uptake Cytotoxicity Assay for acute oral toxicity testing[36]) are provided in the Supplementary Figures, and results for additional assays are available in the GitHub repository described in Section 3.5.

### 4.1 Analysis of Reference Compound Properties Used in Validation Studies

We first examined the properties of reference compound lists used in existing validation studies. These lists are selected by domain experts and reflect practical considerations related to assay purpose, feasibility, and interpretability within specific validation contexts. **Figure 1** compares the structural, physicochemical, and toxicity diversity metrics of the ER-STTA reference compound list with distributions obtained from 10,000 randomly generated compound lists sampled from the corresponding toxicity dataset.

Relative to randomly generated selections, the ER-STTA reference compound list shows lower structural diversity, corresponding to higher internal structural similarity under the present metric definition. This result suggests that, for this assay, the selected compounds occupy a relatively narrower region of structural space than randomly generated lists. Across other validation assays, however, the relative position of the reference list with respect to the random distributions is assay-dependent rather than uniform, as shown in **Supplementary Figure S1**. Representative molecular structures from these reference lists are provided in **Figure 1D and Supplementary Figure S2**.

### 4.2 Genetic Algorithm Optimization Process

To examine the behavior of the GA during optimization, we tracked the evolution of diversity metrics across generations. **Figure 2** shows the progression of population-averaged structural, physicochemical, and toxicity diversity scores over the course of the evolutionary process. Across all objectives, the distributions stabilized over successive generations, consistent with convergence of the search process.

**Figure 3** visualizes the distribution of compound lists in the objective space at the initial and final generations. Compared with the initial random population, compound lists in the final generation occupy distinct regions of the design space, with part of the final population distributed along the Pareto front. A single illustrative Pareto-optimal example was selected from the final Pareto front using the composite score described in the **Supplementary Notes** and used only for illustrative comparisons. Comparable convergence behavior and Pareto front evolution for other assays are shown in **Supplementary Figures S3 and S4**.

### 4.3 Comparison of GA-Derived and Validation Reference Compound Lists

We next compared compound lists generated by the genetic algorithm with the original reference compound lists used in validation studies, as well as with randomly generated compound lists. As shown in **Figures 4A–4C**, GA-derived compound lists occupy regions of the multi-objective design space that differ from those corresponding to both randomly generated and expert-selected reference lists, exhibiting distinct combinations of structural, physicochemical, and toxicity diversity scores.

Visual inspection of an illustrative GA-derived compound list (**Figure 4D**) highlights differences in chemical composition relative to the original validation reference list. Similar distributional patterns across additional validation assays are presented in **Supplementary Figure S5**. These observations support the view that systematic exploration of the multi-objective design space can identify reference compound lists with diversity profiles that differ from those obtained through manual selection or random sampling, without implying preference or superiority among these selections. This illustrative Pareto-optimal example is shown solely to visualize one feasible composition within the trade-off space and is not interpreted as a preferred, recommended, or uniquely meaningful solution.

### 4.4 Illustrative Evaluation Outcomes Associated with Compound List Selection

We next examined how differences in reference compound composition were reflected in observed evaluation outcomes when the same modeling procedure was applied across test sets. To this end, toxicity prediction models were trained under a common protocol, and three types of compound lists were used as held-out test sets: Pareto-front lists identified by the genetic algorithm, intermediate lists sampled during the optimization process, and randomly generated lists.

As shown in **Figure 5**, evaluation metrics differed across these test sets under otherwise comparable training settings. Similar patterns were also observed in other assays (**Supplementary Figure S6**). These differences indicate that reference compound lists occupying different regions of the multi-objective design space correspond to distinct evaluation outcome profiles. Notably, compound lists located closer to the Pareto front showed evaluation outcome profiles that differed from those obtained with randomly generated lists under the same modeling procedure.

These observations should not be interpreted as evidence that Pareto-front lists provide a more rigorous, superior, or preferred evaluation. Rather, they show that observed evaluation outcomes vary with design choices in reference compound selection even when the modeling procedure is held fixed.

## 5 Discussion

In this study, we formulated reference compound selection for toxicity assay validation as an explicit multi-objective design problem. Rather than aiming to improve predictive models or establish more stringent evaluation benchmarks, our primary contribution lies in the formulation itself, which makes the trade-offs underlying reference compound selection explicit and suitable for systematic analysis under controlled assumptions.

A central implication of this formulation is that differences in reference compound selection can be situated and compared as positions within an explicit multi-objective design space, even when the modeling approach and training protocol are held fixed. This makes it possible to relate differences in observed evaluation outcomes to specific regions of the trade-off space in test-set composition, rather than treating them as isolated or ad hoc effects. In this sense, the present study should be understood as a methodological case study on evaluation design across several validation contexts, rather than as a comprehensive survey of all toxicity assay types or a benchmark comparison of predictive models.

This perspective complements prior evaluation design studies that have focused primarily on data-splitting strategies such as random or time-based splits[19–21]. Rather than replacing such protocols, the present formulation addresses a different but complementary question: how the internal composition of reference compound sets structures the evaluation setting itself. For example, emphasizing diversity in toxicological responses may narrow the range of chemical scaffolds represented in a set, whereas prioritizing broad structural coverage can introduce compounds that are less directly aligned with the assay context or endpoint-specific relevance criteria. Reference compound selection and data-splitting strategies therefore operate at different but complementary levels of evaluation design and can be considered jointly in future studies.

Several limitations follow directly from the assumptions made. The objectives considered here capture only three commonly invoked aspects of reference compound selection and do not reflect regulatory definitions or expert judgment in full. In addition, factors such as compound cost, availability, and experimental feasibility were not included because consistent quantitative descriptors were not available. These limitations do not undermine the formulation itself; rather, they define its current scope and clarify which aspects of reference compound selection are being modeled explicitly.

The evaluation analyses raise a distinct interpretive issue. They are not intended to establish that any particular reference compound list provides a more rigorous, superior, or preferred evaluation. Nor do they isolate the effect of test-set composition from all sources of model variability. However, because the same model family, tuning procedure, and evaluation protocol were used throughout, the results support the view that reference compound selection is a non-negligible design factor contributing to variation in observed evaluation outcomes under otherwise comparable modeling conditions. Here, assay-dependent should be understood operationally, as reflecting differences in candidate pools, endpoint definitions, and selection contexts, rather than implying a uniquely identifiable mechanistic explanation.

Overall, this work provides a methodological perspective for treating reference compound selection as an explicit design component in toxicity assay validation that builds on established expert-driven practices. By making underlying trade-offs explicitly comparable within a common design space, the formulation offers a transparent basis for discussing evaluation design choices and their consequences. More broadly, the present results suggest that the structure of the design space is assay-dependent rather than universal, providing a way to examine trade-offs that may otherwise remain implicit in expert-driven selection.

## Author Contribution

Yohei Ohto: Methodology, Software, Investigation, Writing – Original Draft, Visualization.
Yasuhiro Yoshikai: Methodology, Software.
Hiromi Fujimoto: Investigation
Kaoru Sato: Supervision, Writing – Review.
Hiroyuki Kusuhara: Supervision, Writing – Review.
Tadahaya Mizuno: Conceptualization, Resources, Supervision, Project administration, Writing – Original Draft, Writing – Review & Editing, Funding acquisition.

## Conflicts of Interest

The authors declare that they have no conflicts of interest.

## Acknowledgement

We thank all those who contributed to the construction of the datasets employed in the present study such as ICE, TDC, and Tox21. This work was supported by AMED under Grant Number JP22mk0101250h and 23ak0101199h0001. English language editing was assisted by ChatGPT. The final text was reviewed by the authors.

## References

1. Morimoto, B. H., Castelloe, E. & Fox, A. W. Safety pharmacology in drug discovery and development. *Handb. Exp. Pharmacol.* **229**, 65–80 (2015).
2. Van Norman, G. A. Limitations of animal studies for predicting toxicity in clinical trials: Is it time to rethink our current approach? *JACC Basic Transl. Sci.* **4**, 845–854 (2019).
3. Akhtar, A. The flaws and human harms of animal experimentation. *Camb. Q. Healthc. Ethics* **24**, 407–419 (2015).
4. Ahuja, V., Adiga Perdur, G., Aj, Z., Krishnappa, M. & Kandarova, H. In silico phototoxicity prediction of drugs and chemicals by using Derek Nexus and QSAR Toolbox. *Altern. Lab. Anim.* **52**, 195–204 (2024).
5. Thomas, R. S. *et al.* The next generation blueprint of computational toxicology at the U.s. environmental Protection Agency. *Toxicol. Sci.* **169**, 317–332 (2019).
6. Baudy, A. *et al.* Liver microphysiological systems development guidelines for safety risk assessment in the pharmaceutical industry. *Lab Chip* (2019) doi:10.1039/c9lc00768g.
7. Wakefield, I. D., Pollard, C., Redfern, W. S., Hammond, T. G. & Valentin, J.-P. The application of in vitro methods to safety pharmacology. *Fundam. Clin. Pharmacol.* **16**, 209–218 (2002).
8. Marx, U. *et al.* Biology-inspired microphysiological system approaches to solve the prediction dilemma of substance testing. *ALTEX* **33**, 272–321 (2016).
9. Rothfuss, A. *et al.* Collaborative study on fifteen compounds in the rat-liver Comet assay integrated into 2- and 4-week repeat-dose studies. *Mutat. Res.* **702**, 40–69 (2010).
10. Mizumachi, H. *et al.* The inter-laboratory validation study of EpiSensA for predicting skin sensitization potential. *J. Appl. Toxicol.* **44**, 510–525 (2024).
11. Onoue, S. *et al.* Non-animal photosafety assessment approaches for cosmetics based on the photochemical and photobiochemical properties. *Toxicol. In Vitro* **27**, 2316–2324 (2013).
12. United Nations: Economic Commission for Europe. *Globally Harmonized System of Classification and Labelling of Chemicals (GHS).* (United Nations, New York, NY, 2023).
13. Yamaguchi, H., Kojima, H. & Takezawa, T. Predictive performance of the Vitrigel-eye irritancy test method using 118 chemicals: Predictive performance of Vitrigel-eye irritancy test method. *J. Appl. Toxicol.* **36**, 1025–1037 (2016).
14. Sampson J. R. Adaptation in natural and artificial systems (John H. holland). *SIAM Rev. Soc. Ind. Appl. Math.* **18**, 529–530 (1976).
15. Coello Coello, C. A., Lamont, G. B. & van Veldhuizen, D. A. *Evolutionary Algorithms for Solving Multi-Objective Problems.* (Springer, New York, NY, 2007).
16. Sarode, K. & Javaji, S. R. Hybrid Genetic Algorithm and Hill Climbing optimization for the neural network. *arXiv [cs.NE]* (2023).
17. Wang, S. Y. & Tai, K. Structural topology design optimization using Genetic Algorithms with a bit-array representation. *Comput. Methods Appl. Mech. Eng.* **194**, 3749–3770 (2005).
18. Manavi, M., Zhang, Y. & Chen, G. Resource allocation in cloud computing using genetic algorithm and neural network. *arXiv [cs.DC]* (2023).
19. Sheridan, R. P. Time-split cross-validation as a method for estimating the goodness of prospective prediction. *J. Chem. Inf. Model.* **53**, 783–790 (2013).
20. Morita, K., Mizuno, T. & Kusuhara, H. Investigation of a data split strategy involving the time axis in adverse event prediction using machine learning. *J. Chem. Inf. Model.* **62**, 3982–3992 (2022).
21. Landrum, G. A. *et al.* SIMPD: an algorithm for generating simulated time splits for validating machine learning approaches. *J. Cheminform.* **15**, 119 (2023).
22. Zdrazil, B. *et al.* The ChEMBL Database in 2023: a drug discovery platform spanning

multiple bioactivity data types and time periods. *Nucleic Acids Res.* **52**, D1180–D1192 (2024).
23. Top. https://www.jacvam.go.jp/.
24. Bell, S. *et al.* An integrated chemical environment with tools for chemical safety testing. *Toxicol. In Vitro* **67**, 104916 (2020).
25. Huang, K. *et al.* Therapeutics Data Commons: Machine learning datasets and tasks for drug discovery and development. *arXiv [cs.LG]* (2021).
26. Richard, A. M. *et al.* The Tox21 10K compound library: Collaborative chemistry advancing toxicology. *Chem. Res. Toxicol.* **34**, 189–216 (2021).
27. Kuhn, M., Letunic, I., Jensen, L. J. & Bork, P. The SIDER database of drugs and side effects. *Nucleic Acids Res.* **44**, D1075-9 (2016).
28. Kim, S. *et al.* PubChem 2025 update. *Nucleic Acids Res.* **53**, D1516–D1525 (2025).
29. Landrum, G. *et al. Rdkit/Rdkit: 2025_03_2 (Q1 2025) Release*. (Zenodo, 2025). doi:10.5281/ZENODO.15286010.
30. Deb, K., Pratap, A., Agarwal, S. & Meyarivan, T. A fast and elitist multiobjective genetic algorithm: NSGA-II. *IEEE Trans. Evol. Comput.* **6**, 182–197 (2002).
31. Rogers, D. & Hahn, M. Extended-connectivity fingerprints. *J. Chem. Inf. Model.* **50**, 742–754 (2010).
32. Tehrany, E. A., Fournier, F. & Desobry, S. Simple method to calculate octanol–water partition coefficient of organic compounds. *J. Food Eng.* **64**, 315–320 (2004).
33. Ertl, P., Rohde, B. & Selzer, P. Fast calculation of molecular polar surface area as a sum of fragment-based contributions and its application to the prediction of drug transport properties. *J. Med. Chem.* **43**, 3714–3717 (2000).
34. Volume 16: How to Detect and Handle Outliers. *Google Books* https://books.google.com/books/about/Volume_16_How_to_Detect_and_Handle_Outli.html?hl=ja&id=FuuiEAAAQBAJ.
35. Preprint at https://www.jacvam.go.jp/files/news/20160513_2.pdf.
36. Prieto, P. M. D. P., Griesinger, C., Amcoff, S. P. & Whelan, M. *EURL ECVAM Recommendation on the 3T3 Neutral Red Uptake Cytotoxicity Assay for Acute Oral Toxicity Testing*. (2013).
37. Fortin, F.-A., Rainville, F., Gardner, M.-A., Parizeau, M. & Gagné, C. DEAP: evolutionary algorithms made easy. *J. Mach. Learn. Res.* **13**, 2171–2175 (2012).
38. Chen, T. & Guestrin, C. XGBoost: A Scalable Tree Boosting System. *arXiv [cs.LG]* (2016) doi:10.1145/2939672.2939785.
39. van der Burg, B. *et al.* Optimization and prevalidation of the in vitro AR CALUX method to test androgenic and antiandrogenic activity of compounds. *Reprod. Toxicol.* **30**, 18–24 (2010).

## Figures and Tables

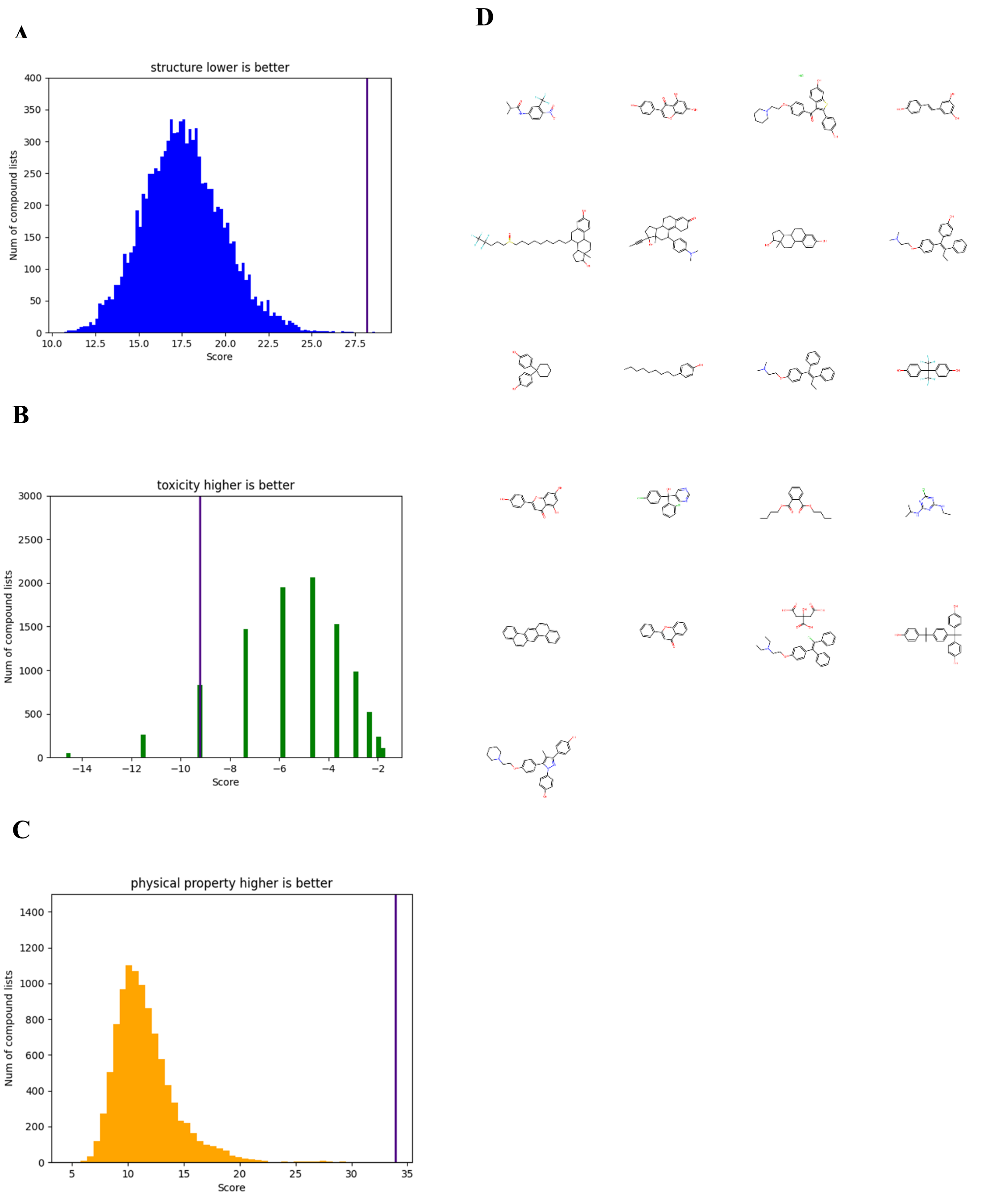

**Figure 1. Characteristics of Reference Compound Lists Based on Objective Function Scores.**
(A–C) Histograms illustrating the distribution of diversity scores (structural, physicochemical, and toxicity diversity) from 10,000 randomly generated compound lists. A vertical purple line indicates the corresponding diversity score for the reference compound list used in the validation study.
(D) Chemical structures of compounds included in the reference compound list.

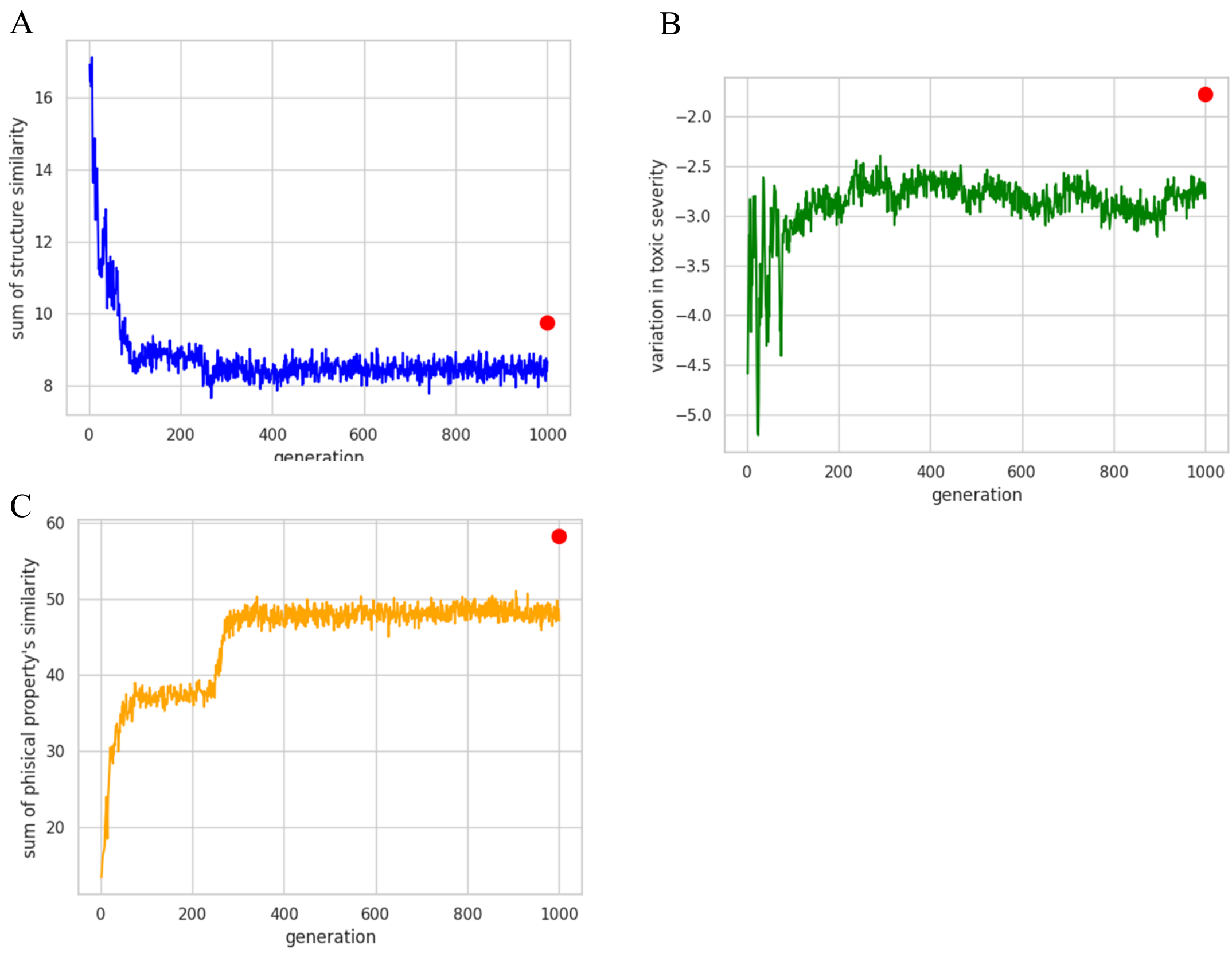

**Figure 2. Convergence of the Genetic Algorithm.**

(A–C) Plots showing changes in average population scores across generations for structural diversity (A), toxicity diversity (B), and physicochemical diversity (C). The distributions of each diversity metric stabilize over successive generations, indicating convergence of the evolutionary search process. Red dots indicate the scores of one illustrative Pareto-optimal example (selected using the composite score described in the Supplementary Notes) from the final generation's Pareto front.

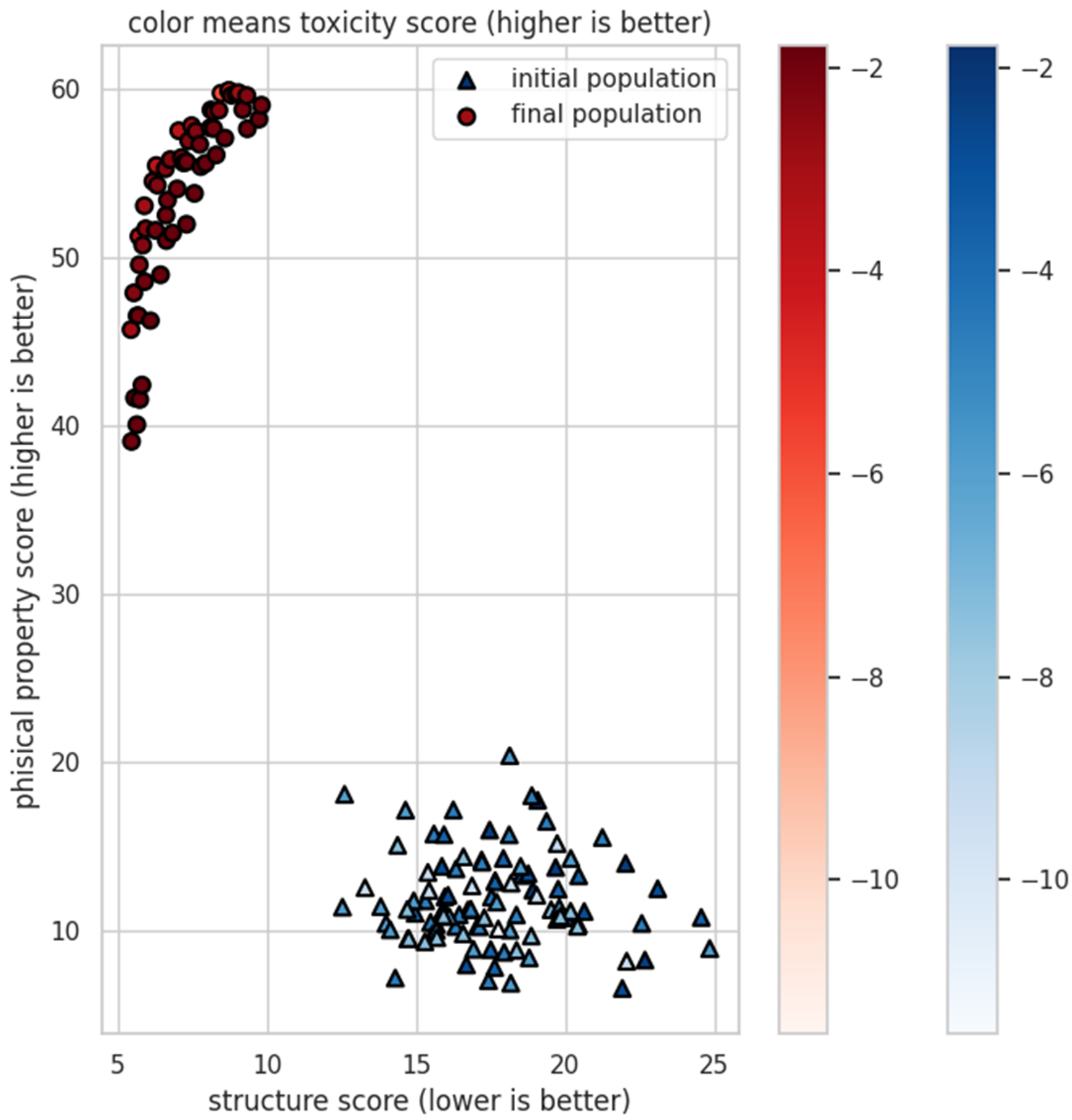

**Figure 3. Evolution of the Pareto Front from Initial to Final Generation.**

Scatter plot illustrating the distribution of compound lists in the multi-objective design space. Triangular points represent compound lists from the initial generation, while circular points indicate compound lists from the final generation. The x- and y-axes denote structural diversity score (lower values indicate lower similarity) and physicochemical property diversity score (higher values indicate greater dispersion), respectively. Color indicates the toxicity diversity score (higher values indicate greater diversity).

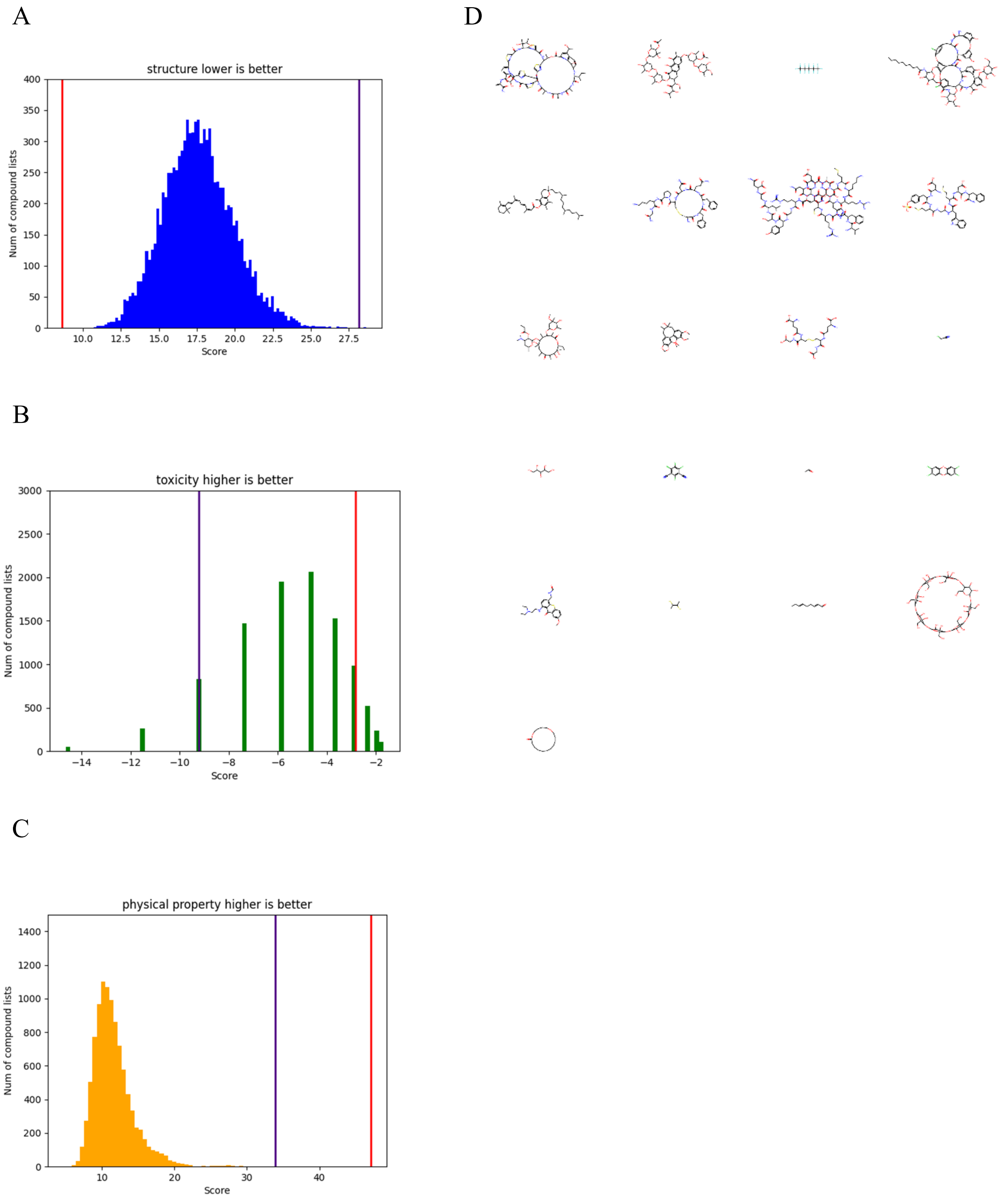

**Figure 4. Comparison of Objective Function Scores for GA-Derived, Validation Reference, and Randomly Generated Compound Lists.**

(A–C) Histograms displaying distributions of diversity scores (structural, physicochemical, and toxicity diversity) from 10,000 randomly generated compound lists. Vertical lines indicate the scores of the reference compound list used in the validation study (purple) and one illustrative Pareto-optimal example selected from the Pareto front identified by the genetic algorithm (red).

(D) Chemical structures of compounds in the illustrative GA-derived compound list.

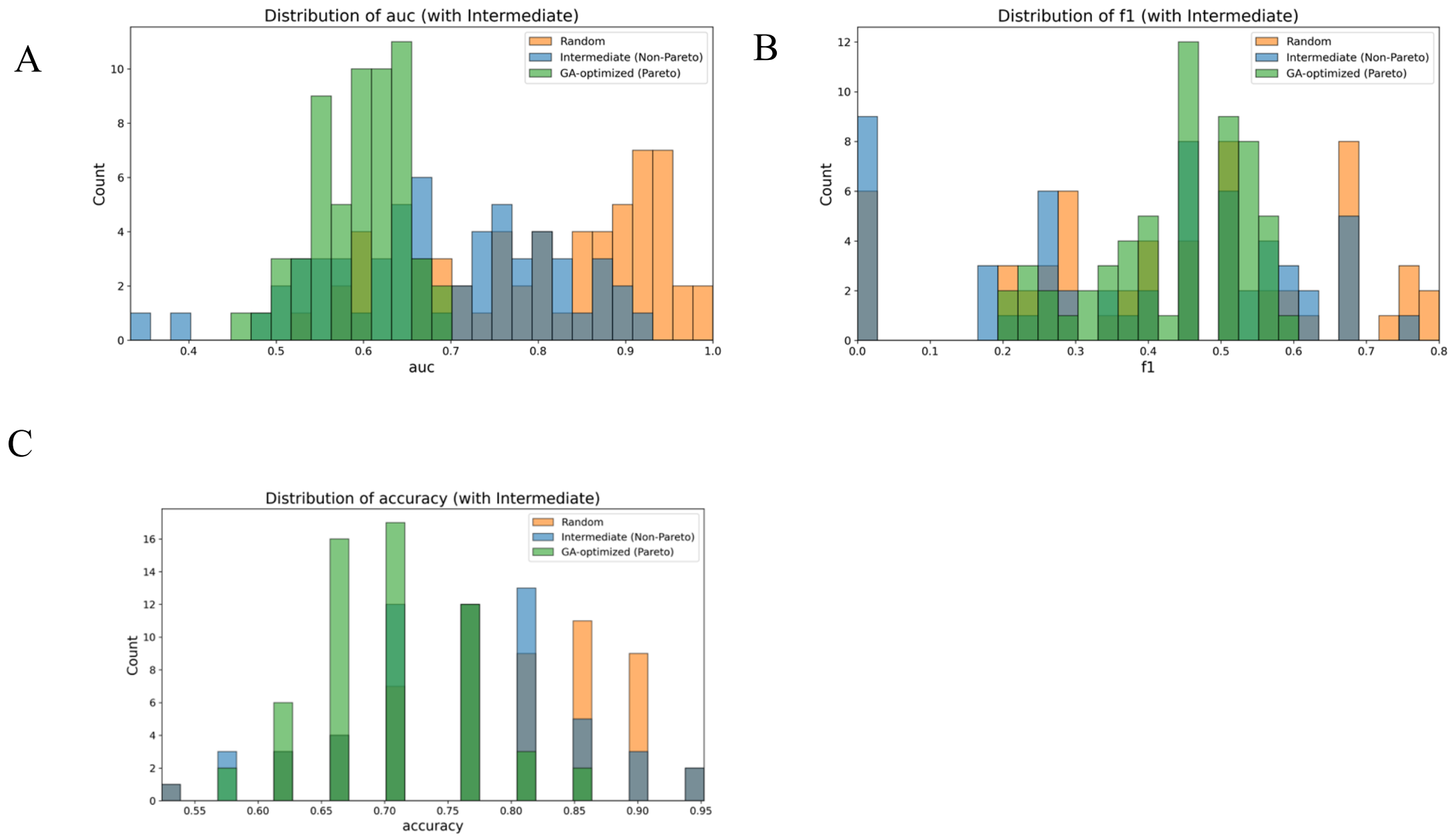

**Figure 5. Illustrative Evaluation Outcomes Associated with Reference Compound Selection.**

Toxicity prediction results obtained under a common modeling protocol in the 09_02 antagonist assay using three types of held-out reference compound lists. Green histograms represent compound lists sampled from the Pareto front identified by the genetic algorithm, blue histograms correspond to intermediate compound lists sampled during the optimization process, and orange histograms denote randomly generated compound lists. Panels A–C show AUC, F1, and accuracy scores, respectively.

**Table 1. Validation studies used for optimization of reference compound lists.** Full details, including compound numbers and data linkages (ICE/TDC/Tox21), are provided in **Supplementary Table 1**.

| **Category** | **JaCVAM Test No.** | **Assay Name** | **Endpoints / Type** | **Primary Data Source** |
|---|---|---|---|---|
| Acute toxicity | 07_02 | 3T3 Neutral Red Uptake (NRU) Cytotoxicity Assay | Cell viability (NRU); supports identification of substances not classified for acute oral toxicity | ICE |
| Endocrine disruption | 09_01 | VM7Luc ER TA | ER transactivation (agonist/antagonist) | TDC / Tox21 |
| | 09_02 | ER STTA | ER transactivation (agonist/antagonist) | TDC / Tox21 |
| | 09_04 | AR-EcoScreen | AR transactivation (agonist/antagonist) | TDC / Tox21 |
| | 09_05 | AR-CALUX | AR transactivation (agonist/antagonist) | TDC / Tox21 |
| | 09_07 | hrER binding (OECD TG 493) | ERα binding affinity (radioligand/competition) | TDC / Tox21 |
| Developmental toxicity | 10_01 | Embryonic Stem Cell Test (EST) | mES differentiation inhibition + cytotoxicity | ICE |
| | 10_02 | Hand1-Luc EST | Hand1-reporter–based differentiation toxicity + cytotoxicity | ICE |

# Supporting Information for "Exploring Multi-Objective Trade-offs in Reference Compound Selection for Validation Studies of Toxicity Assays"

Yohei Ohto[1], Yasuhiro Yoshikai[1], Hiromi Fujimoto[1], Kaoru Sato[2], Hiroyuki Kusuhara[1], Tadahaya Mizuno[1, 3†]

1 Laboratory of Molecular Pharmacokinetics, Graduate School of Pharmaceutical Sciences, The University of Tokyo, 7-3-1 Hongo, Bunkyo, Tokyo, Japan
2 National Institute of Health Sciences, 3-25-26 Tonomachi, Kawasaki-ku, Kawasaki City, Kanagawa Prefecture, Japan
3 The Institute of Statistical Mathematics (ISM), Research Organization of Information and Systems, Tachikawa, Tokyo, 190-8562, Japan.
† Author to whom correspondence should be addressed

# Supplementary Notes

## Note S1. Methodological Limitations and Future Extensions

Several practical factors and modeling choices were not incorporated into the current formulation. While these simplifications enabled direct comparison with existing validation studies under controlled assumptions, they also define the present scope of the analysis.

### Size of Compound Lists

In this study, compound list size was fixed to match that used in each original validation assay, ensuring a consistent basis for comparison within each validation context. However, this constraint prevents exploration of formulations in which list size itself is treated as a design variable.

Allowing variable list sizes could support analysis of more complex design questions, such as how coverage of chemical space, assay relevance, and experimental feasibility trade off against one another. Future extensions could therefore include list size as an additional decision variable and examine how it alters the structure of the design space.

### Cost and Experimental Feasibility

Factors such as compound cost, handling feasibility, and inter-laboratory reproducibility are important in actual validation studies but were not included here because consistent quantitative descriptors were not available in structured form.

Validation assays often prioritize compounds that are stable, easy to handle, and likely to result in reproducible measurements across laboratories. Although these factors are relevant to practical implementation, they are rarely represented in a form that can be incorporated directly into the present formulation. If such metadata become available in the future, they could be introduced as additional objectives or constraints, thereby extending the realism of the design space without changing the basic formulation presented here.

## Note S2. Genetic Algorithm Implementation Details

Reference compound selection was explored using a multi-objective genetic algorithm (GA) implemented with the DEAP Python library (version 1.4.2)[37]. The NSGA-II (Non-dominated Sorting Genetic Algorithm II)[30] framework was adopted to maintain a diverse set of non-dominated solutions in the design space.

Each candidate reference compound list was represented as an individual consisting of a fixed number of compounds, matching the size of the corresponding JaCVAM validation study. The initial population consisted of 100 compound lists randomly sampled from the candidate compound pool defined for each validation context.

Evolutionary operators included crossover and mutation. Crossover exchanged subsets of compounds between two parent lists, whereas mutation replaced one or more compounds with alternatives drawn from the remaining candidate pool. The crossover rate and mutation rate were set to 0.8 and 0.2, respectively.

Selection was performed using the NSGA-II procedure. At each generation, individuals were ranked by Pareto dominance, and crowding distance was used to promote diversity along the Pareto front. The evolutionary process was repeated for 1,000 generations. At the final generation, the Pareto front was extracted using the *tools.ParetoFront()* function implemented in DEAP.

The GA parameters, including population size, number of generations, crossover rate, and mutation rate, were not optimized for performance. Instead, standard values commonly used in NSGA-II–based exploratory analyses were adopted to support stable exploration and reproducible sampling of the design space. Accordingly, these parameters were treated as fixed implementation choices rather than variables of interest in the present study.

## Note S3. Definitions and Normalization of Diversity Metrics Structural Diversity

For every pair of compounds within a list, the Tanimoto index of their ECFP4 fingerprints[31] was calculated. The sum of these pairwise Tanimoto indices was then used as the structural diversity metric.

$$\sum_{i=1}^{n}\sum_{k=1}^{i-1} Tanimoto\ similarity(ECFP(c_i),\ ECFP(c_k)) \rightarrow min.$$

Minimizing the sum of pairwise Tanimoto similarities therefore corresponds to reducing overall structural redundancy within the compound list and increasing coverage of distinct chemical scaffolds.

### Physicochemical Property Diversity

For each compound, the LogP value[32] and the TPSA value[33] were calculated. For all compound pairs, these descriptors were standardized using the robust z-score method[34], and the Euclidean distance in the standardized property space was computed:

$$\sum_{i=1}^{n}\sum_{k=1}^{i-1} D(c_i, c_k) \rightarrow max$$

$$D(c_i, c_k) = \left\| \begin{pmatrix} robust_z(logp(c_i)) - robust_z(logp(c_k)) \\ robust_z(TPSA(c_i)) - robust_z(TPSA(c_k)) \end{pmatrix} \right\|_2 .$$

Larger cumulative distances in the standardized property space indicate greater dispersion of physicochemical characteristics across compounds, reflecting increased physicochemical diversity within the list.

### Toxicity Diversity

Toxicity diversity within a compound list was evaluated in an assay-specific manner.

When toxicity values were expressed as continuous variables (e.g., $LD_{50}$ for acute toxicity), diversity was quantified using the coefficient of variation (CV) of log-transformed toxicity values within each compound list:

$$CV(\log(TOX\ SCORE(c_0)),\ \cdots,\ \log(TOX\ SCORE(c_n))) \rightarrow max\ \ (if\ TOX\ SCORE\ is\ continuous).$$

Higher CV indicates broader coverage across potency ranges within the assay-specific response scale.

When toxicity endpoints were binary, diversity was quantified using the log-likelihood of a binomial distribution with success probability fixed at 0.5, which assigns higher scores to compound lists with more balanced class compositions. For a list with $n_{positive}$ toxic and $n_{negative}$ non-toxic compounds (total $n = n_{positive} + n_{negative}$), the diversity score is calculated as:

$$\log\binom{n}{n_{positive}} + n_{positive} log\,(0.5) + (n - n_{positive}) log\,(0.5) \rightarrow \max\ (if\ TOX\ SCORE\ is\ 0\ or\ 1).$$

This formulation assigns higher scores to compound lists with more balanced class compositions, thereby capturing diversity in terms of response category representation without assuming a specific class ratio.

## Note S4. Composite Score for Representative List Selection

To facilitate visualization and illustrative comparison, one illustrative Pareto-optimal example was selected from the Pareto front using a composite score defined as:

$$\{Composite\ Score\} = 10 \cdot \{Structural\ Diversity\} + 2 \cdot \{Physicochemical\ Property\ Diversity\} + 1 \cdot \{Toxicity\ Diversity\}.$$

This composite score was not used during the optimization process and does not affect the identification or structure of the Pareto front. Its sole purpose is to select one example from the set of Pareto-optimal solutions for presentation.

The weights were chosen as an interpretable heuristic reflecting the commonly emphasized role of chemical structure coverage in reference compound selection for toxicity assay validation, while still accounting for physicochemical and toxicity diversity. Alternative weighting schemes would select different illustrative examples without altering the Pareto front itself or the conclusions regarding the underlying trade-off structure.

## Note S5. Evaluation Procedures for Illustrative Analyses

### Evaluation Based on Objective Function Scores

Compound lists generated by the genetic algorithm were compared with randomly generated compound lists and with the original reference lists used in validation studies. Comparisons were conducted using the same three diversity metrics employed during optimization: structural diversity, physicochemical property diversity, and toxicity diversity. These comparisons characterize how different selection strategies are distributed within the multi-objective design space.

### Evaluation of Model Performance Characteristics

To examine illustrative evaluation outcomes, XGBoost[38] models (version 2.1.4) were trained on datasets excluding the compounds in each test list and evaluated on the corresponding held-out compounds. For each training–test split, model hyperparameters were optimized using the same Optuna-based procedure to ensure consistent model fitting across different compound lists. XGBoost was selected as a representative and widely used machine learning method for toxicity prediction on tabular chemical descriptors, serving here as a stable evaluation probe rather than as a subject of model comparison. Hyperparameter optimization was performed using Optuna over the following parameters: max_depth (3–12), learning_rate (0.01–0.2), n_estimators (50–1000), subsample (0.6–1.0), colsample_bytree (0.6–1.0), gamma (0–1), and min_child_weight (1–10), with the objective function fixed to squared error regression.

Prediction metrics such as accuracy and AUC were compared across test sets. Because each test set consists of a distinct set of compounds, differences in these metrics were interpreted as reflecting variation associated with test-set composition under a common modeling procedure, while not excluding contributions from model fitting variability or limited test-set size. Accordingly, these analyses are not intended to isolate pure test-set effects or to compare modeling strategies, but to examine whether observed evaluation outcomes vary systematically with reference compound selection under otherwise comparable conditions.

## Supplementary Figures and Tables

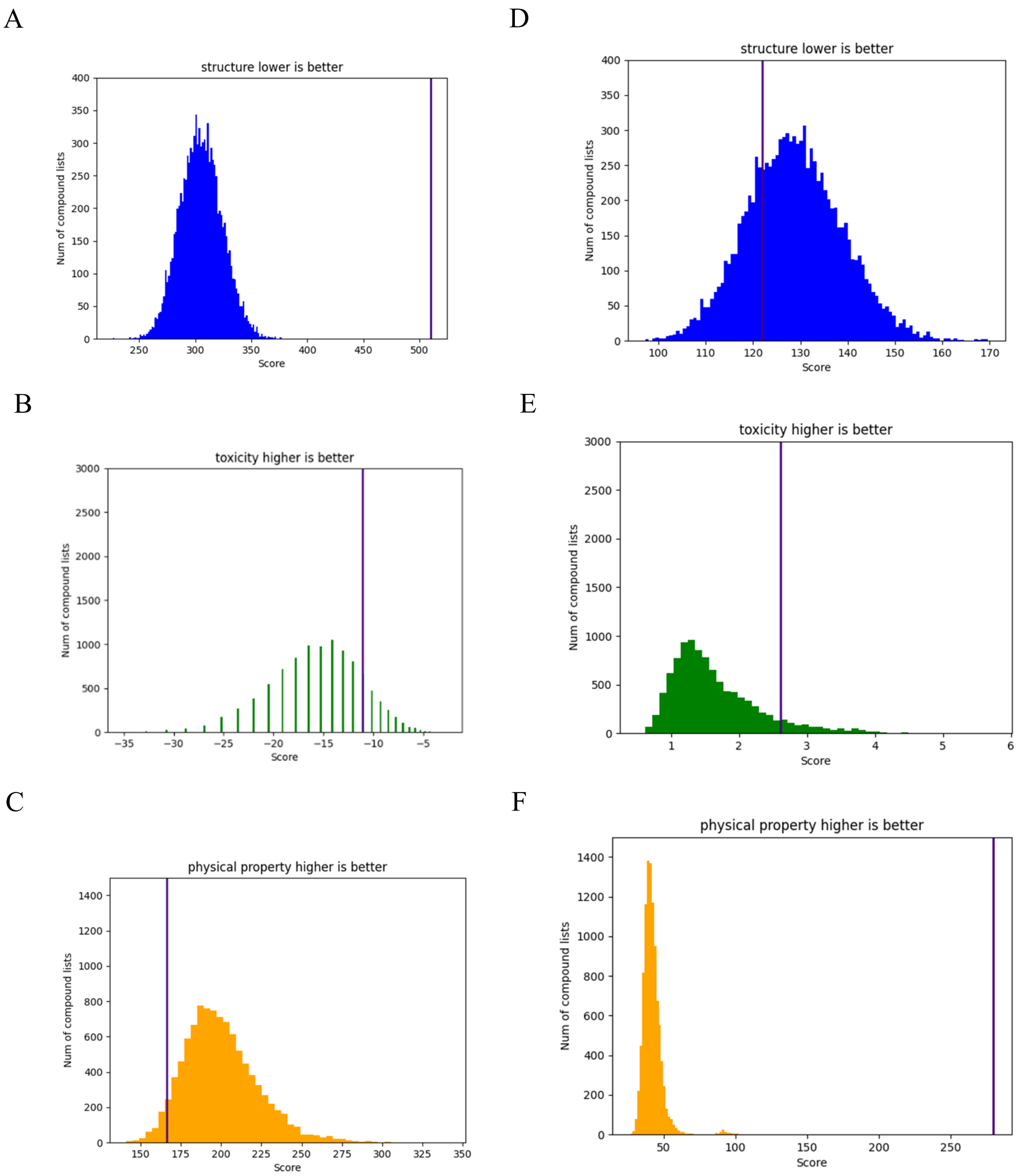

**Supplementary Figure 1. Characteristics of Validation Compound Lists Based on Objective Function Scores.**

Histograms illustrating the distribution of diversity scores (structural, physicochemical, and toxicity diversity) from 10,000 randomly generated compound lists. A vertical blue line indicates the corresponding diversity score of the reference compound list used in the validation study. Panels (A–C) show results for the 09_02 agonist assay, and panels (D–F) show results for the 07_02 assay.

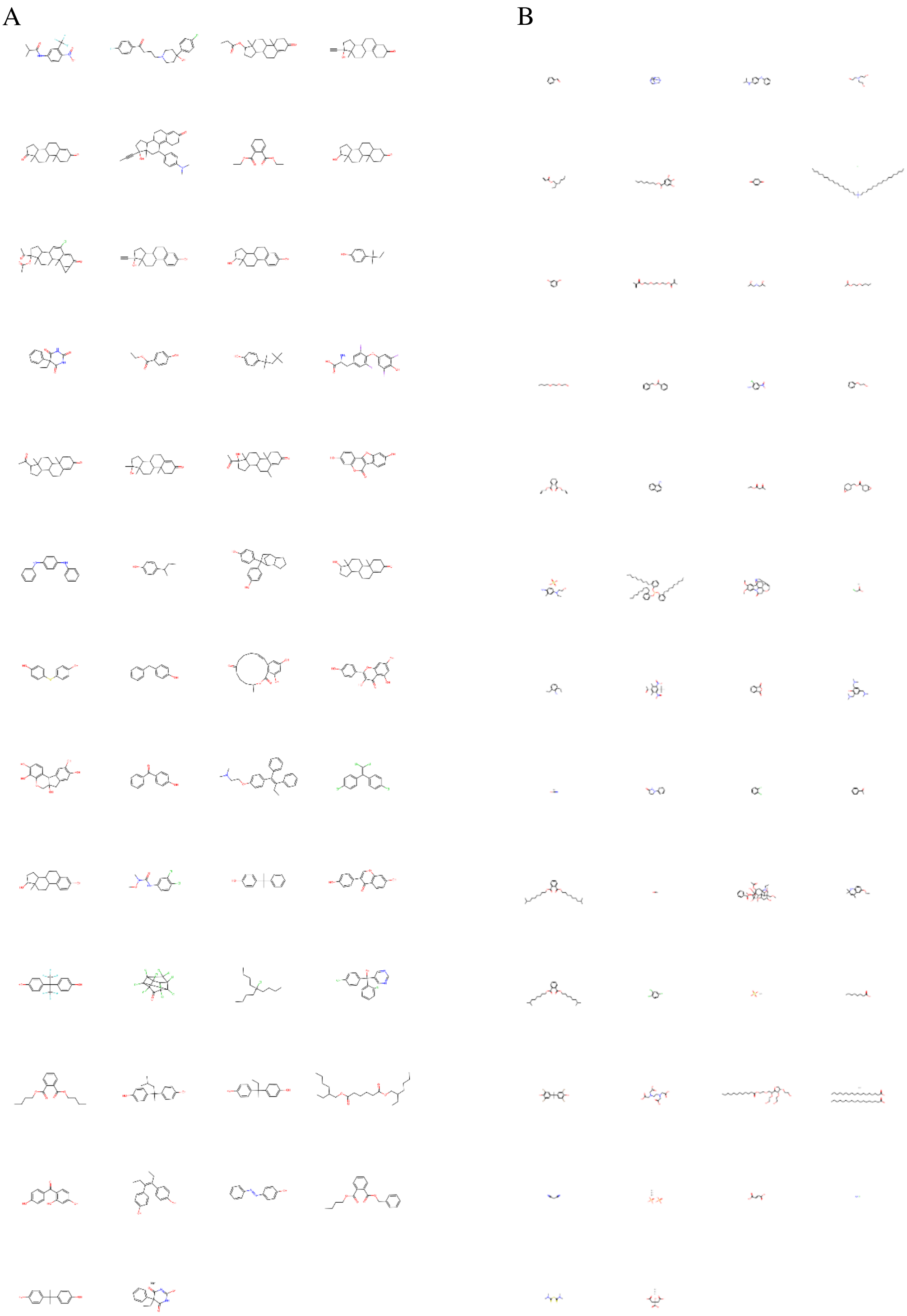

**Supplementary Figure 2. Structures of Reference Compound Lists.**

Representative chemical structures of compounds included in the reference compound lists. Panel (A) shows the 09_02 agonist assay, and panel (B) shows the 07_02 assay.

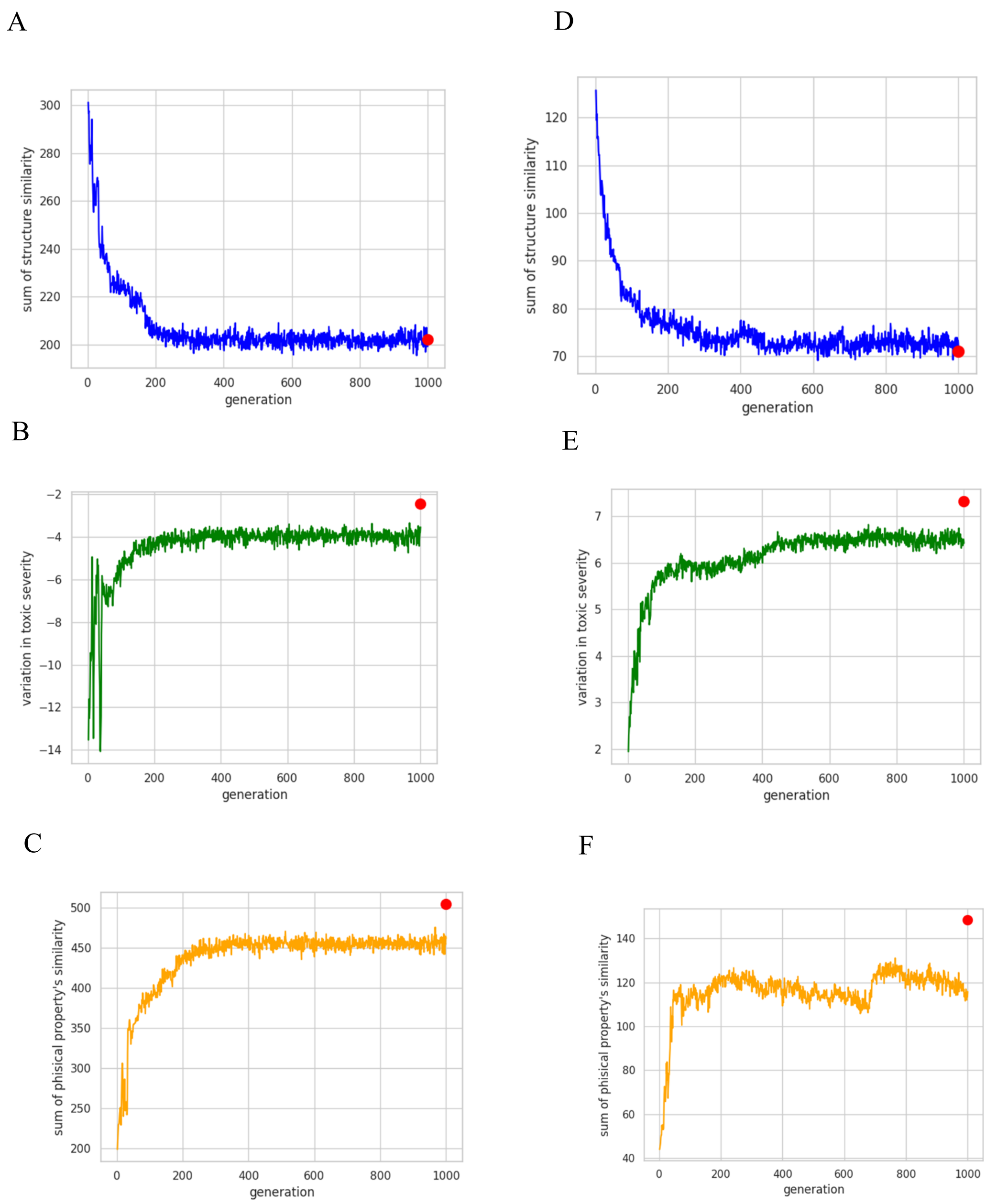

**Supplementary Figure 3. Convergence of the Genetic Algorithm in Additional Assays.**

Plots showing changes in average population scores across generations for structural diversity (A (09_02 agonist), D (07_02)), toxicity diversity (B (09_02 agonist), E (07_02)), and physicochemical diversity (C (09_02 agonist), F (07_02)). The distributions of each diversity metric stabilize over successive generations, indicating convergence of the evolutionary search process. Red dots indicate the scores of one illustrative Pareto-optimal example (selected using the composite score described in the Supplementary Notes) from the final generation's Pareto front.

A B

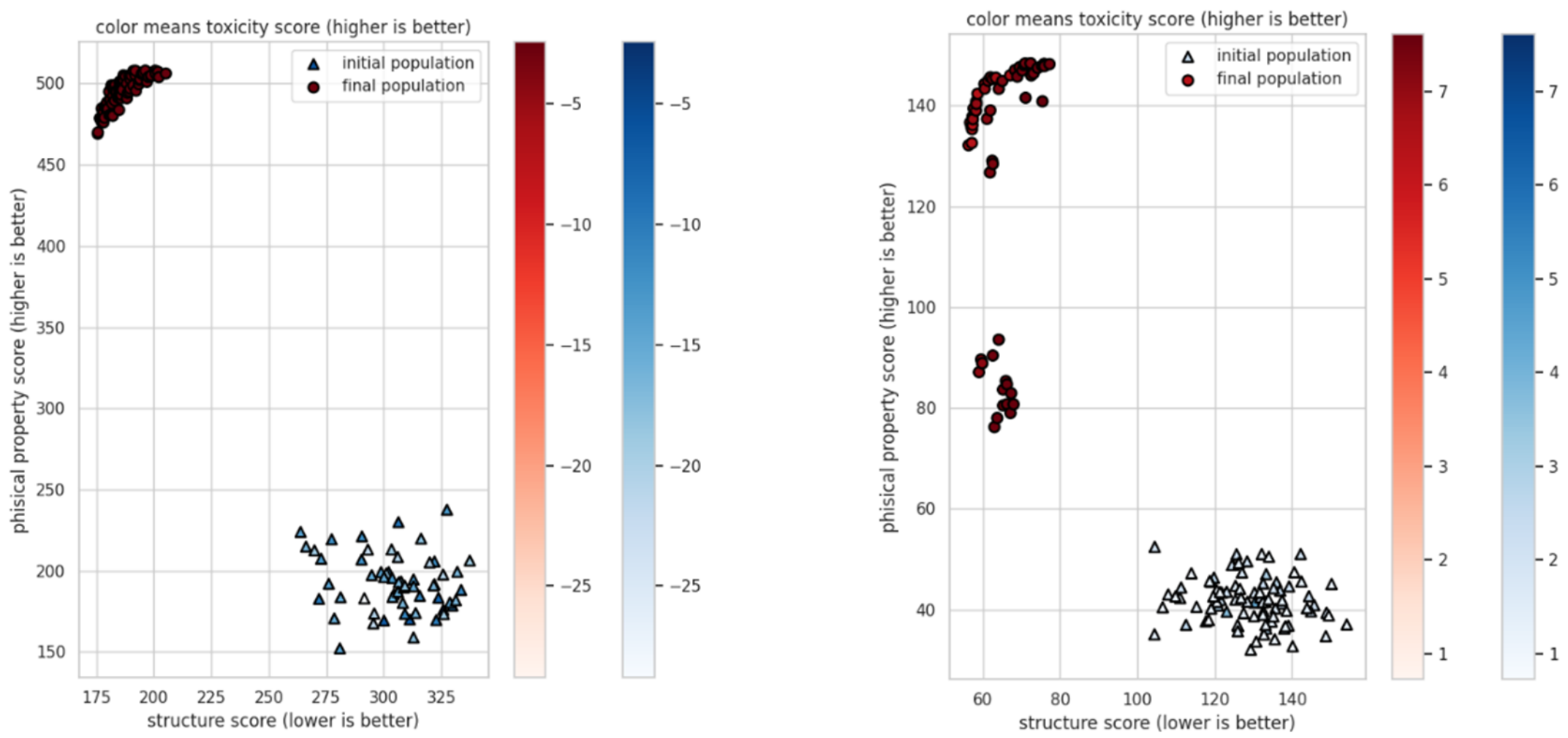

**Supplementary Figure 4. Distribution of Compound Lists in the Objective Space Across Generations.**

Scatter plot illustrating the distribution of compound lists in the multi-objective design space. Triangular points represent compound lists from the initial generation, while circular points indicate compound lists from the final generation. The x- and y-axes denote structural diversity score (lower values indicate lower similarity) and physicochemical property diversity score (higher values indicate greater dispersion), respectively. Color indicates the toxicity diversity score (higher values indicate greater diversity).

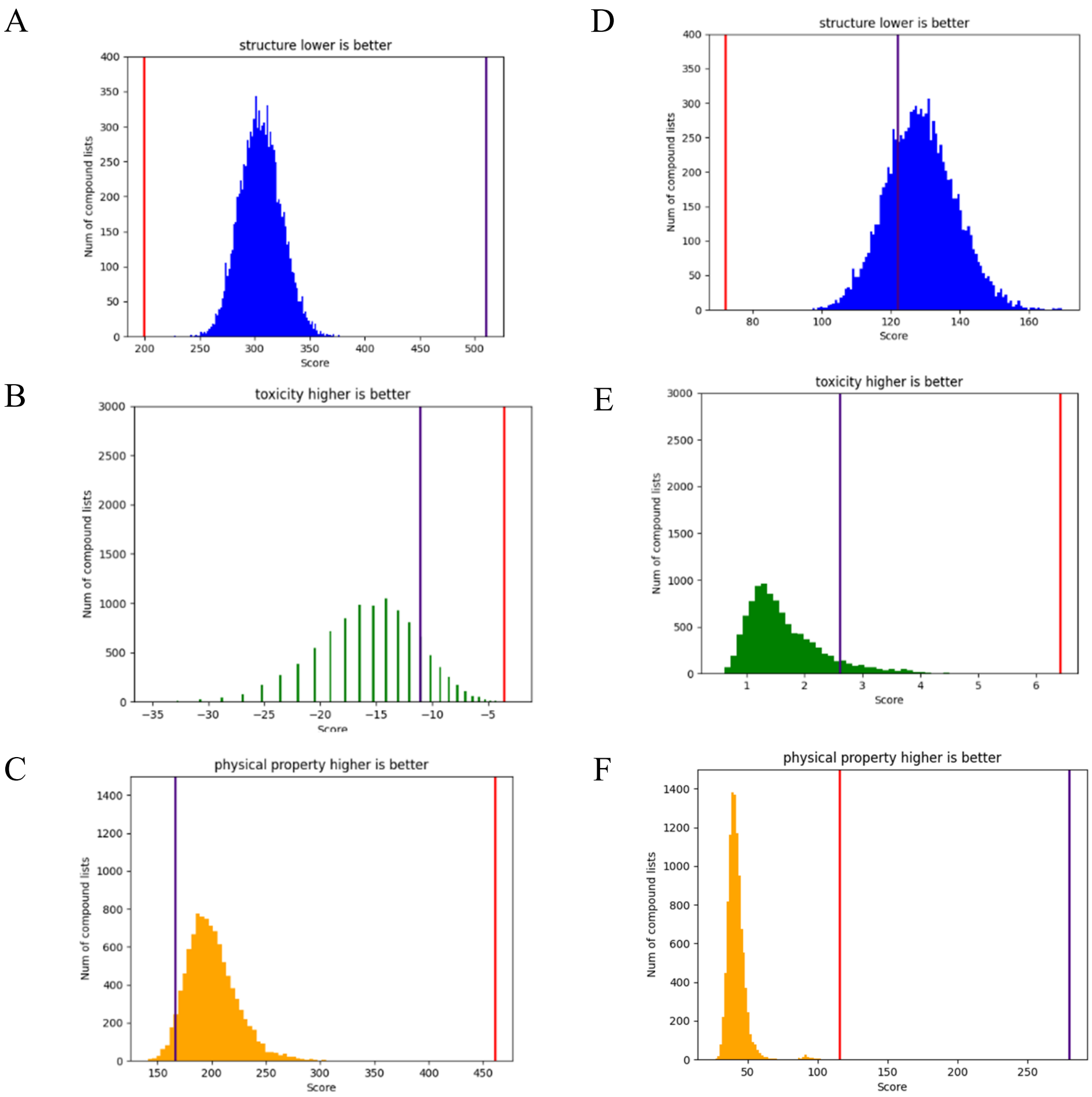

**Supplementary Figure 5. Comparison of Objective Function Scores for GA-Derived, Validation Reference, and Randomly Generated Compound Lists in Additional Assays.**

(A–C) Histograms showing distributions of diversity scores (structural, physicochemical, and toxicity diversity) for the 09_02 agonist assay, and (D–F) corresponding distributions for the 07_02 assay, based on 10,000 randomly generated compound lists. Vertical lines indicate the scores of the reference compound list used in the validation study (purple) and one illustrative Pareto-optimal example selected from the Pareto front identified by the genetic algorithm (red).

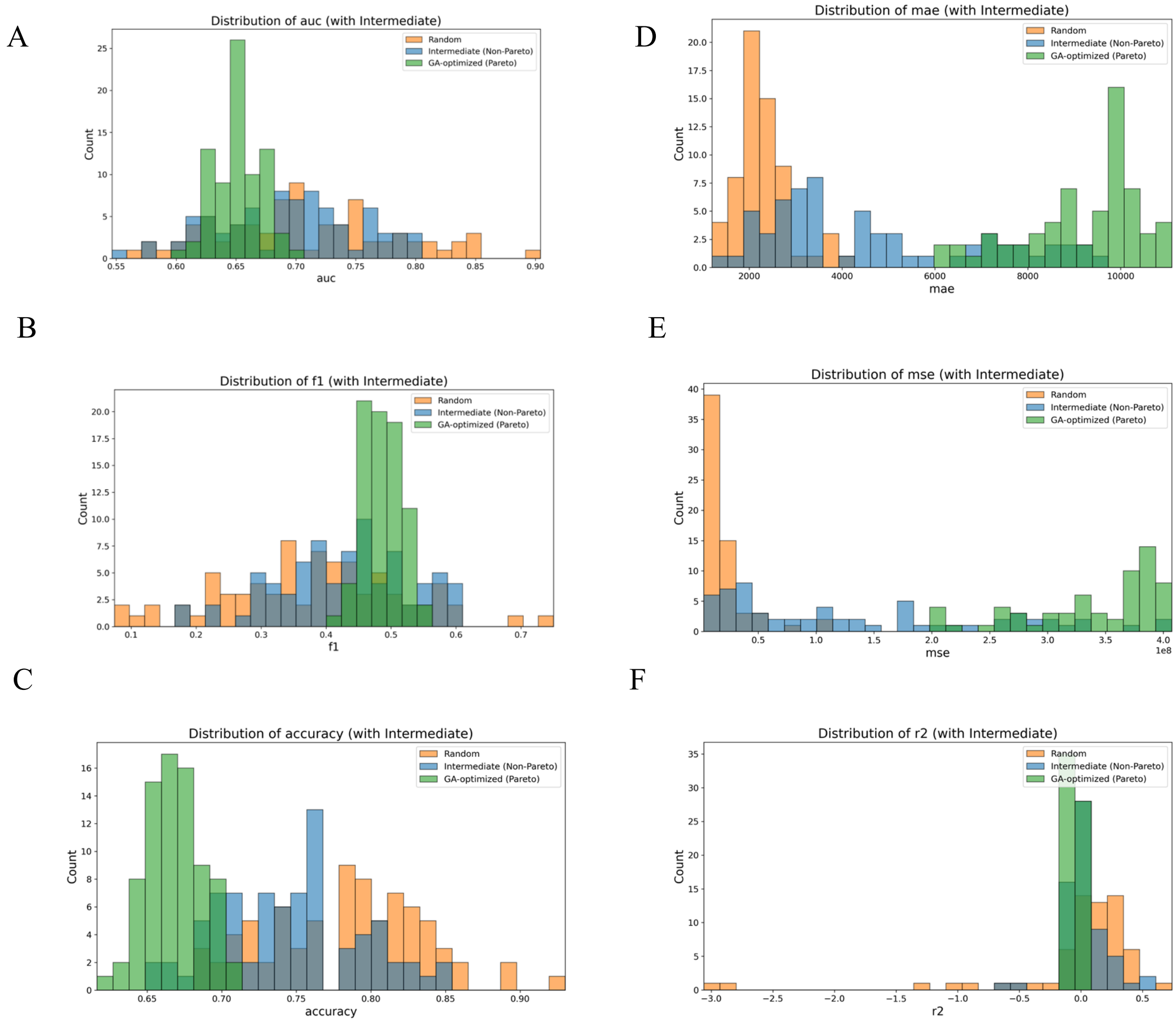

**Supplementary Figure 6. Illustrative Evaluation Outcomes Associated with Reference Compound Selection in Additional Assays.**

Toxicity prediction results obtained under the same modeling framework as in Figure 5 using three types of held-out reference compound lists. Green histograms represent compound lists sampled from the Pareto front identified by the genetic algorithm, blue histograms correspond to intermediate compound lists sampled during the optimization process, and orange histograms denote randomly generated compound lists. For the 09_02 agonist assay, panels A–C show AUC, F1, and accuracy scores, respectively. For the 07_02 assay, panels D–F show mean absolute error (MAE), mean squared error (MSE), and R-squared ($R^2$) scores.

## Supplementary Table 1. Detailed List of Validation Studies for Toxicity Testing Used in This Study.

| JaCVAM Test Number | JaCVAM Test Name | Number of Compounds | | compound dataset |
|---|---|---|---|---|
| 07_acute toxicity | 01_Cytotoxicity Testing | 72 | Complete DL file set: page 33 (Data from JaCVAM's unpublished proposal) | ICE Acute Oral Toxicity |
| 07_acute toxicity | 02_Cytotoxicity Testing | 56 | Complete DL file set: page 38 (Data from JaCVAM's unpublished proposal) | ICE Acute Oral Toxicity |
| 09_endocrine disruptors | 01_VM7 Luc ER TA assay | 42 (agonist) | Complete DL file set: page 26 (Data from JaCVAM's unpublished proposal) | TDC Use the appropriate tox21 test from among these |
| | | 25 (antagonist) | Complete DL file set: page 28 (Data from JaCVAM's unpublished proposal) | TDC Use the appropriate tox21 test from among these |
| 09_endocrine disruptors | 02_ER-STTA assay | 86 (agonist) | Complete DL file set: page 32-34 (Data from JaCVAM's unpublished proposal) | TDC Use the appropriate tox21 test from among these |
| | | 21 (antagonist) | Complete DL file set: page 35 (Data from JaCVAM's unpublished proposal) | TDC Use the appropriate tox21 test from among these |
| 09_endocrine disruptors | 04_AR-Ecoscreen | 10 (agonist) | Complete DL file set: page 29, 31 (Data from JaCVAM's unpublished proposal) | TDC Use the appropriate tox21 test from among these |
| | | 10 (antagonist) | Complete DL file set: page 30, 32 (Data from JaCVAM's unpublished proposal) | TDC Use the appropriate tox21 test from among these |
| 09_endocrine disruptors | 05_AR-CALUX | 11 (agonist) | Bart et al., Reprod.[39] Toxicol., 2010 43: table 1 | TDC Use the appropriate tox21 test from among these |
| | | 9 (antagonist) | Bart et al., Reprod.[39] Toxicol., 2010 43: table 2 | TDC Use the appropriate tox21 test from among these |
| 09_endocrine disruptors | 07_hrER in vitro study | 36 | Complete DL file set: page 28 (Data from JaCVAM's unpublished proposal) | TDC Use the appropriate tox21 test from among these<br>TDC Use the appropriate tox21 test from among these |
| 10. Developmental Toxicity Prediction Test | 01_Embryonic Stem Cell Technology (EST) | 18 | Complete DL file set: page 4 (Data from JaCVAM's unpublished proposal) | ICE DART |
| 10. Developmental Toxicity Prediction Test | 02_Hand1-Luc EST | 16 | Complete DL file set: page 56, 69 (Data from JaCVAM's unpublished proposal) | ICE DART |